\documentclass[12pt]{article}

\usepackage{epsfig}
\usepackage{amsfonts}
\usepackage{amssymb}
\usepackage{amsmath}   
\usepackage{amsthm}
\usepackage{latexsym}

\newtheorem{theorem}{Theorem}[section]

\newtheorem{corollary}[theorem]{Corollary}

\newtheorem{remark}[theorem]{Remark}

\DeclareSymbolFont{AMSb}{U}{msb}{m}{n}
\DeclareMathSymbol{\N}{\mathbin}{AMSb}{"4E}
\DeclareMathSymbol{\Z}{\mathbin}{AMSb}{"5A}
\DeclareMathSymbol{\R}{\mathbin}{AMSb}{"52}
\DeclareMathSymbol{\Q}{\mathbin}{AMSb}{"51}
\DeclareMathSymbol{\I}{\mathbin}{AMSb}{"49}
\DeclareMathSymbol{\C}{\mathbin}{AMSb}{"43}

\title{Particle Filters for Multiscale Diffusions}

\author{Anastasia Papavasiliou\footnote{Department of Statistics, University of Warwick, Coventry, CV4 7AL, UK. Email: {\tt a.papavasiliou@warwick.ac.uk}. The author has been partially supported by a Marie Curie International Reintegration Grant, MIRG-CT-2005-029160.}}

\begin{document}
\maketitle

\begin{abstract}
We consider multiscale stochastic systems that are partially observed at discrete points of the slow time scale. We introduce a particle filter that takes advantage of the multiscale structure of the system to efficiently approximate the optimal filter.
\end{abstract}



\section{Introduction}

We are interested in the problem of estimating a function of a multiscale process that can be approximated by a diffusion which lives in the slow scale, when it is partially observed. Such problems come up in many applications, such as molecular dynamics, climate modelling or estimation of stochastic volatility using agent-based models (see \cite{Givon-Kupferman-Stuart} for general discussion of multiscale models and \cite{E-Liu-VandenEijnden2} and \cite{Majda-Timofeyev-VandenEijnden} for applications to kinetic Monte-Carlo and climate modelling respectively).

In this paper, we focus on the problem of estimating the slow component of a continuous multiscale process from partial and discrete observations of it. More specifically, we have an ${\mathbb R}^{p+q}$ process $X^\epsilon=(X^\epsilon_t)_{t\geq 0} = (X_t^{(1,\epsilon)},X_t^{(2,\epsilon)})_{t\geq 0}$ that satisfies the following multiscale stochastic differential equation:

\begin{equation}
\label{full X}
\left\{
\begin{array}{ccccc}
dX_t^{(1,\epsilon)} &=& a(X_t^{(1,\epsilon)},X_t^{(2,\epsilon)}) dt &+& \sigma_1(X_t^{(1,\epsilon)},X_t^{(2,\epsilon)}) dW^{(1)}_t \\
dX_t^{(2,\epsilon)} &=& \frac{1}{\epsilon} b(X_t^{(1,\epsilon)},X_t^{(2,\epsilon)}) dt &+& \frac{1}{\sqrt{\epsilon}}\sigma_2(X_t^{(1,\epsilon)},X_t^{(2,\epsilon)}) dW^{(2)}_t
\end{array}
\right.
\end{equation}
where $X^{(1,\epsilon)}_t\in{\mathbb R}^{p}$, $X^{(2,\epsilon)}_t\in{\mathbb R}^{q}$ and $W^{(1)}_t$ and $W^{(2)}_t$ are two independent Wiener processes in ${\mathbb R}^p$ and ${\mathbb R}^q$ respectively. Let $\mu$ be the initial distribution, i.e. $\mu = {\mathcal L}(X^\epsilon_0)$. We denote by $\mu_1$ and $\mu_2$ the marginals on $X^{(1,\epsilon)}_0$ and $X^{(2,\epsilon)}_0$ respectively. 

We observe the process $(X^\epsilon_t)_{t\geq 0}$ through $Y^\epsilon = (Y^\epsilon_{k\Delta })_{k = 0,\dots,T}$, where $\Delta  \sim {\mathcal O}(1)$, i.e. the observations live in the the same scale as $X_t^{(1,\epsilon)}$, which we call the {\it slow time scale}. In fact, let us assume for simplicity that $\Delta = 1$. The process $Y^\epsilon$ is given by
\begin{equation}
	\label{discrete Y}
		Y^\epsilon_{k} = h(X^{(1,\epsilon)}_{k}, v_k),
\end{equation}
where $(v_k)_k$ are i.i.d. random variables with known distribution.

Our goal is to compute the conditional distribution of the slow process $X^{(1,\epsilon)}_t$ given the observations, at observation points $k=1,\dots,T$, or, equivalently, compute the expectations
\begin{equation}
\label{pi}
	\pi^\epsilon_k(f) := {\mathbb E}\left( f(X^{(1,\epsilon)}_{k }) | Y^\epsilon_{\ell}, \ell=1,\dots,k \right)
\end{equation}
for all continuous bounded functions $f$ on ${\mathbb R}^p$, i.e. $f\in{\mathcal C}_b({\mathbb R}^p)$, and $k=0,\dots,T$.

In \cite{DelMoral-Jacod-Protter}, the authors discuss this problem for an arbitrary diffusion $(X_t)_{t\geq 0}$ and they develop a particle filter that approximates the conditional distribution (also called {\it optimal filter}). The additional difficulty compared to discrete systems is that of simulating the process $(X_t)_t$ between the observation points, i.e. simulate $X_{(k+1)}$ given $X_{k}$. In \cite{DelMoral-Jacod-Protter}, this is done by applying the Euler discretization scheme. The step size is given as a function of the number of particles used by the particle filter, chosen so as to optimize the convergence rate. An alternative approach has been recently suggested in \cite{Fearnhead-Papaspiliopoulos-Roberts}.

In the case of multiscale diffusions, both the Euler discretization scheme and the MCMC method described in \cite{Fearnhead-Papaspiliopoulos-Roberts} become inefficient. However, if we are only interested in the slow scale marginal of the optimal filter given by (\ref{pi}), we can avoid these problems by replacing the multiscale diffusion by the approximation of the slow scale constructed by applying the averaging principle. When the averaged equation is not available in closed form, we construct a further approximation of its drift and variance using short simulations of the multiscale system (see \cite{E-Liu-VandenEijnden, Gear-Kevrekidis-Theodoropoulos}).

In section \ref{review}, we review some of the basic results of \cite{DelMoral-Jacod-Protter} for discretely and partially observed diffusions. In section \ref{averaged}, we describe the algorithm and analyze the approximation error in the case were the averaged equation is available in closed form. In section \ref{hmm}, we do the same for the case were the averaged equation is not available in closed form and we apply the heterogeneous multiscale method to approximate it. Finally, in section 4, we discuss how to extent this approach to continuous observation processes.

\section{Discretely and partially observed diffusions: a review}
\label{review}

Suppose that $X = (X_t)_{t\geq 0}$, with $X_t \in {\mathbb R}^d$, is a diffusion of the form 
\begin{equation}
\label{diffusion}
dX_t = a(X_t)dt + \sigma(X_t)dW_t,
\end{equation}
with initial distribution $\mu = {\mathcal L}(X_0)$. The diffusion process is observed through
\begin{equation}
\label{obs 1}
Y_{k} = h(X_{k},v_k),
\end{equation}
where $(v_k)_{k\geq 0}$ are i.i.d. random variables, such that the conditional probability admits a density $g$, i.e. ${\mathbb P}(Y_{k}\in dy | X_{k } = x) = g(x,y)dy$, and $g$ is bounded and explicitly known. We want to approximate the optimal filter
\begin{equation}
\label{full filter}
\pi_k = {\mathbb P}(X_{k}|Y_{\ell },\ell=1,\dots,k).
\end{equation}
In \cite{DelMoral-Jacod-Protter}, the authors approximate (\ref{full filter}) using a combination of the Euler method and the discrete particle filter. The exact algorithm is as follows:
\begin{itemize}
	\item{{\it Initialization (k=0):} Simulate $N$ independent random variables $(\xi_0^j)_{j=1}^N$ from the initial distribution $\mu$.}
	\item{For $k>0$
	\begin{enumerate}
		\item{{\it Evolution}: Simulate 
		\[\hat{\xi}^j_{k}\sim p^{(N)}_1(\xi^j_{k-1},\cdot),\] 
		where $p^{(N)}_1(x,\cdot)$ is the forward Euler approximation with step $\frac{1}{\sqrt{N}}$ of the transition kernel 
		\[p_1(x,\cdot) = {\mathbb P}\left(X_1\in\cdot|X_0 = x\right).\] }
		\item{{\it Resampling}: Simulate $N$ new random variables $(\xi^j_{k})_{j=1}^N$ from 
		\[ \xi^j_{k} \sim \frac{1}{N}\sum_{j=1}^N \frac{w^j_{k}}{\sum_{i=1}^M w^i_k} \delta_{\hat{\xi}^j_{k}}, \] 
		where the weights $(w^j_{k})_{j=1}^N$ are the likelihood of observing $Y_{k}$ if $X_{k} = \hat{\xi}^j_{k}$, i.e.  $w^j_{k} := g(\hat{\xi}^j_{k},Y_{k})$.}
	\end{enumerate}}
\end{itemize}
Then, the particle filter $\pi^N_k = \frac{1}{N}\sum_{j=1}^N \delta_{\xi^j_k}$ converges weakly to the optimal filter $\pi_k$ defined in (\ref{full filter}). More precisely, the following holds:

\begin{theorem}[Del Moral -- Jacod -- Protter, \cite{DelMoral-Jacod-Protter}]
\label{thm 1}
For all bounded Borel functions $f\in{\mathcal B}_b({\mathbb R}^d)$, all $k=0,\dots,T$ and all $N\geq 0$, the approximation error will be bounded by
\begin{equation}
\label{approximation error 1}
{\mathbb E}\left| \pi^N_k(f) - \pi_k(f) \right| \leq \frac{C_k}{\sqrt{N}} \| f \|_\infty,
\end{equation}
under the following assumptions
\begin{enumerate}
	\item{The functions $a(\cdot)$ and $\sigma(\cdot)$ are two times differentiable with bounded derivatives of all orders up to two.}
	\item{The covariance matrix is uniformly non degenerate, i.e. $\sigma\sigma^{t}(\cdot)>\eta>0$. }
\end{enumerate}
The constant $C_k$ depends on the drift and variance of the diffusion, the likelihood function $g$ and $k$.
\end{theorem}

If the likelihood function $g(x,y)$ is bounded above and below, i.e. there exists a constant $K$ such that $\frac{1}{K}\leq g(x,y) \leq K$ for all $x$ and $y$, then the constant $C_k$ in theorem \ref{thm 1} takes the following form: 
\[ C_k = (2+2\alpha)\frac{(8K^{2T})^{k+1}-8K^{2T}}{8K^{2T}-1}, \]
where $\alpha$ is such that 
\[ \sup_x | p^{(N)}_1 f(x) - p_1 f(x)| \leq \frac{\alpha}{\sqrt{N}}\|f\|_\infty,\]
with $p^{(N)}_1$ and $p_1$ as above. So, $\alpha$ is the constant that appears in the upper bound of the error of the approximation of the distribution of $X_1$ by the forward Euler method (see \cite{Bally-Talay}). Consequently, if we apply the Euler discretization method to the multiscale system (\ref{full X}), the constant $\alpha$ will be of order $\alpha\sim{\mathcal O}(\frac{1}{\epsilon})$. 

\begin{corollary}
Under the assumptions of theorem \ref{thm 1}, if the diffusion process and the observations are of the form (\ref{full X}) and (\ref{discrete Y}) respectively,  then the error of the slow scale marginal of the particle filter described above becomes
\begin{equation}
\label{approximation error 2}
{\mathbb E}\left| \pi^{\epsilon,N}_k(f) - \pi^\epsilon_k(f) \right| \leq \frac{C^\prime_k}{\epsilon\sqrt{N}} \| f \|_\infty,
\end{equation}
\end{corollary}
The above corollary shows that if the diffusion process that we want to estimate is a multiscale diffusion, the particle filter described in \cite{DelMoral-Jacod-Protter} will no longer be efficient, just as the Euler discretization method will not be efficient. 

\section{The multiscale case}
\label{averaged}

Since the observations live in the slow scale, we can only hope to get a good approximation of the slow scale marginal of the optimal filter and, consequently, we focus on the approximation of $\pi^\epsilon_k(f)$ given by (\ref{pi}). A quite natural thing to do in order to avoid simulating the whole multiscale process -- which, as we have already seen, is problematic -- is to try and replace the slow scale process $(X^{1,\epsilon}_t)_t$ by a diffusion $(\bar{X}_t)_t$ in ${\mathbb R}^p$ that does not depend on the fast scale process $(X^{2,\epsilon}_t)_t$. This is, indeed, possible under the following assumption: $\exists \lambda>0$ such that $\forall x_1 \in {\mathbb R}^p$ and $\forall x_2, x^\prime_2 \in{\mathbb R}^q$,
\begin{equation}
\label{mixing assumption}
<x_2 - x^\prime_2, b(x_1,x_2)-b(x_1,x^\prime_2)> + \|\sigma_2(x_1,x_2)-\sigma_2(x_1,x^\prime_2) \|^2 \leq -\lambda |x_2 - x^\prime_2|^2, 
\end{equation}
where $<\cdot,\cdot>$, $|\cdot|$ and $\|\cdot\|$ denote the Euclidean inner product and norm in ${\mathbb R}^q$ and the Frobenius norm in ${\mathbb R}^q\times {\mathbb R}^q$, respectively. In other words, we require both $b$ and $\sigma$ to grow sublinearly. This assumption implies that if we fix $X^{1,\epsilon}_t \equiv x_1$ in (\ref{full X}), $X^{2,\epsilon}_t$ converges to its unique invariant distribution $\nu_{x_1}$ exponentially fast, with rate $\frac{\lambda}{\epsilon}$. In fact, the necessary assumption is not (\ref{mixing assumption}) but this exponential ergodicity property. We approximate the process $(X^{1,\epsilon}_t)_t$ by the diffusion process $(\bar{X}_t)_t$ satisfying
\begin{equation}
\label{averaged SDE}
d\bar{X}_t = \bar{a}(\bar{X}_t)dt + \bar{\sigma}(\bar{X}_t)dW_t,\ \ \bar{X}_0 \sim \mu_1 = {\mathcal L}(X^{1,\epsilon}_0),
\end{equation}
where 
\begin{equation}
\label{new drift}
\bar{a}(x) = \int_{{\mathbb R}^q} a(x,z) \nu_x(dz)
\end{equation}
and 
\begin{equation}
\label{new std}
\bar{\sigma}(x) = \left( \int_{{\mathbb R}^q} \sigma_1(x,z)^2 \nu_{x}(dz)  \right)^\frac{1}{2}
\end{equation}
From now on, let us assume that the assumptions of theorem \ref{thm 1} and (\ref{mixing assumption}) hold. Then, it is a well-known result, often referred to as the {\it averaging principle}, that $X^{1,\epsilon}_t \rightarrow \bar{X}_t$ as $\epsilon \rightarrow 0$. More specifically, the following holds (see \cite{Freidlin-Wentzell}):
\begin{equation}
\label{weak averaging}
\sup_{0\leq t\leq T}\left| {\mathbb E}f(X^{1,\epsilon}_t) - {\mathbb E}f(\bar{X}_t) \right| \leq C_{f,T}\cdot \epsilon,\ \forall f\in{\mathcal C}_b({\mathbb R}^p). 
\end{equation}
This estimate suggests that we can approximate $\pi^\epsilon_k(f)$ given by (\ref{pi}) by $\bar{\pi}_k(f)$ defined by
\begin{equation}
\label{bar pi}
\bar{\pi}_k(f) := {\mathbb E}\left( f(\bar{X}_k)\ |\ \bar{Y}_\ell = Y^\epsilon_\ell, \ell = 1,\dots,k \right), 
\end{equation}
where $\bar{Y}_k = h(\bar{X}_k,v_k)$ and $(v_k)_k$ are i.i.d. random variables as in (\ref{discrete Y}). Indeed, it is a straight forward consequence of (\ref{weak averaging}) and Proposition 2.1 of \cite{DelMoral-Jacod-Protter} that
\[ {\mathbb E}\left| \pi^\epsilon_k(f) - \bar{\pi}_k(f) \right| \leq C_1 \epsilon \|f\|_\infty.\]
If we cannot compute $\bar{\pi}_k$ explicitly, we approximate it by the particle filter $\bar{\pi}^N_k$ described in section \ref{review}. Then, the total error will be bounded by
\begin{equation}
\label{total error 1}
{\mathbb E}\left| \pi^\epsilon_k(f) - \bar{\pi}^N_k(f) \right| \leq C \left( \epsilon + \frac{1}{\sqrt{N}} \right)\|f\|_\infty.
\end{equation}
Thus, if $\epsilon$ is small, it is much more efficient to approximate $\pi^\epsilon_k$ by $\bar{\pi}^N_k$ rather than $\pi^{\epsilon,N}_k$ in (\ref{approximation error 2}), i.e. if we are willing to accept an approximation error of order $\delta$, we will, in general, achieve this with a much smaller number of simulations (and computing time) if we compute $\bar{\pi}^N_k$ rather than $\pi^{\epsilon,N}_k$. 

\section{Approximating the averaged equation}
\label{hmm}

In the previous section, we argued that it is, in general, more efficient to approximate the slow marginal of the optimal filter $\pi^\epsilon_k$ by replacing the slow component of multiscale diffusion by another diffusion, which we call {\it averaged diffusion}, and then applying the particle filter algorithm, rather than applying it directly to the multiscale diffusion. However, in order to simulate the averaged diffusion $(\bar{X}_t)_t$ that replaces the slow scale process $(X^{1,\epsilon}_t)_t$, we need to know its drift and its standard deviation given by (\ref{new drift}) and (\ref{new std}) respectively. In most cases, these are not going to be explicitly known. Then, we replace (\ref{new drift}) and (\ref{new std}) by their Monte Carlo estimates, as in \cite{E-Liu-VandenEijnden}.

First, we define a new family of diffusion processes as follows. For each $x\in{\mathbb R}^p$, we define the process $Z_t(x)$ as the solution of the following stochastic differential equation:
\begin{equation}
\label{Z}
dZ_t(x) = b(x,Z_t(x)) dt + \sigma_2(x,Z_t(x)) dV_t,\ Z_0(x)\sim\mu_2,
\end{equation}
where $V_t$ is an ${\mathbb R}^q$-valued Wiener process. We also define a new approximation to the transition kernel $\bar{p}_1(x,\cdot)$, where $\bar{p}_t(x,\cdot) := {\mathbb P}\left( \bar{X}_t\in\cdot | \bar{X}_0 = x \right)$, so that we can simulate from it exactly, as follows:

\begin{enumerate}
			\item{ For $k = 0$:
			
			\noindent Simulate $M$ independent random variables $(\zeta^{i}_{k,n})_{i=1}^M$ from the forward Euler approximation to the distribution of $Z_{n}(x)$ defined in (\ref{Z}), with step $\delta t$ and initial distribution $\mu_2$.	Simulate $\xi_1$ from 
			\[ \xi_1 \sim {\rm Gsn}\left( \Delta t\left(\frac{1}{M} \sum_{i=1}^M a(x, \zeta^{i}_{k,n})) \right) , \Delta t\left( \frac{1}{M} \sum_{i=1}^M \sigma_1(x, \zeta^{i}_{k,n})^2 \right)\right),\]
			where we denote by ${\rm Gsn}(\mu,\tau^2)$ the Gaussian distribution with mean $\mu$ and variance $\tau^2$. Note that we implicitly assume that $q=1$, in order to simplify notation.}
			\item{ For $k=1,\dots,\left\lfloor \frac{1}{\Delta t}\right\rfloor-1$:
			
			\noindent For all $i=1,\dots,M$, set $\zeta^{i}_{k,0} = \zeta^{i}_{k-1,n}$  and simulate $\zeta^{i}_{k,n}$ from the forward Euler approximation to the transition kernel ${\mathbb P}\left( Z_{n}(\xi_k)\in\cdot | Z_0(\xi_k)= \zeta^{i}_{k,0} \right)$ with step $\delta t$. Then, simulate $\xi_{k+1}$ from 
			\[ \xi_{k+1} \sim {\rm Gsn}\left( \Delta t\left(\frac{1}{M} \sum_{i=1}^M a(\xi_k, \zeta^{i}_{k,n})) \right) , \Delta t\left( \frac{1}{M} \sum_{i=1}^M \sigma_1(\xi_k, \zeta^{i}_{k,n})^2 \right)\right).\]}
			\item{ For $k = \left\lfloor \frac{1}{\Delta t}\right\rfloor$:
			
			\noindent As in the previous step, set $\zeta^{i}_{k,0} = \zeta^{i}_{k-1,n}$  and simulate $\zeta^{i}_{k,n}$ from the forward Euler approximation to the transition kernel ${\mathbb P}\left( Z_{n}(\xi_k)\in\cdot | Z_0(\xi_k)= \zeta^{i}_{k,0} \right)$ with step $\delta t$. Then, simulate $\tilde{X}_1$ from 
				\[ \tilde{X}_1 \sim {\rm Gsn}\left( (1-k\Delta t)\left(\frac{1}{M} \sum_{i=1}^M a(\xi_k, \zeta^{i}_{k,n})) \right) , (1-k\Delta t)\left( \frac{1}{M} \sum_{i=1}^M \sigma_1(\xi_k, \zeta^{i}_{k,n})^2 \right)\right).\]}
\end{enumerate}

One can extent the weak convergence theorem in \cite{E-Liu-VandenEijnden} for $\sigma_1 \neq 0$, to get an estimate of the approximation error of the transition kernel. More specifically,
\begin{equation}
\label{weak error}
|{\mathbb E}f(\tilde{X}_1)-{\mathbb E}f(\bar{X}_1)| \leq C_f\left( \Delta t + \delta t + \frac{e^{-\frac{1}{2}\lambda n}}{1 - e^{-\frac{1}{2}\lambda n}}(\Delta t + \Delta t^2) + \frac{\Delta t}{M}\right)
\end{equation}
Let us now define a new particle filter, similar to the one in section \ref{review}, only the evolution of the particles between observation points follows the algorithm above, for $\delta t = \Delta t = \frac{1}{\sqrt{N}}$ and $n=M=1$. The choice $M=1$ might seem surprising at first, but actually gives optimal bounds (see \cite{E-Liu-VandenEijnden}, section 2.4). The reason is that the Monte-Carlo estimation is done by averaging both in time and independent realizations but averaging in time also improves the initialization error. So, it is in theory preferable to average one long path rather than many short ones.

Let us name this new particle filter $\tilde{\pi}^N_k$. Notice that for these values of $\Delta t,\delta t,M,n$, (\ref{weak error}) becomes
\begin{equation*}
|{\mathbb E}f(\tilde{X}_1)-{\mathbb E}f(\bar{X}_1)| \leq C^\prime_f\frac{1}{\sqrt{N}}
\end{equation*}
Then, the total approximation error becomes
\begin{equation}
\label{total error 2}
{\mathbb E}\left| \pi^\epsilon_k(f) - \tilde{\pi}^N_k(f) \right| \leq C \left( \epsilon + \frac{1}{\sqrt{N}} \right)\|f\|_\infty.
\end{equation}

To study the efficiency of this particle filter, suppose that we want to achieve a total error of order ${\mathcal O}(\epsilon)$. Then, if we apply the particle filter algorithm of section \ref{review} to the multiscale system, the number of simulations needed will be of order ${\mathcal O}(\frac{1}{\epsilon^6}(p+q))$: at each step, we simulate $N\sqrt{N}(p+q)$ random variables -- we need $\sqrt{N}(p+q)$ simulations for the evolution of each particle and we have $N$ particles -- and we need $N\sim{\mathcal O}(\frac{1}{\epsilon^4})$, since the total error is given by (\ref{approximation error 2}). 

On the other hand, if we approximate the optimal filter by $\tilde{\pi}^N_k$, we need $N\sim{\mathcal O}(\frac{1}{\epsilon^2})$ to get a total error of order ${\mathcal O}(\epsilon)$. For this particle filter, the number of random variables we simulate at each step is $N(\sqrt{N}p+Nq)$ -- $N$ is the number of particles and we simulate $\sqrt{N}p + N q$ random variables for the evolution of each particle. Notice that since $M=1$, we estimate the drift and variance of (\ref{averaged SDE}) using the final value of only one path of the appropriate process $Z_t(\cdot)$. The reason why we discard the rest of the path is to allow the distribution of $Z_t(\cdot)$ to get close to the invariant distribution of the process. Consequently, we need a total of ${\mathcal O}(\frac{1}{\epsilon^3}p+\frac{1}{\epsilon^4}q)$ simulations, which shows that we can, indeed, achieve substantial improvement in the efficiency of the algorithm by replacing the multiscale system by the averaged diffusion, even when this not explicitly known.

\begin{remark}
 In order to approximate the drift and variance of the averaged process, we need to be able to simulate random variables from the invariant distributions $\nu_x$, for the appropriate $x$. We do that by simulating the process $Z_t(x)$, whose distribution converges exponentially fast to the invariant distribution $\nu_x$. Notice, however, that if $x$ and $x^\prime$ are close, the distributions $\nu_{x}$ and $\nu_{x^\prime}$ will also be close as a result of the smoothness of the drift and variance. Thus, we can improve the efficiency of the algorithm further by correlating the simulations of $Z_t(x)$ and $Z_t(x^\prime)$ as in \cite{Papavasiliou-Kevrekidis} or by using the simulations of one process to initialize the other.
\end{remark}

\section{Conclusions}

This analysis can also be applied for more general observation processes. For example, suppose that we observe $(Y^\epsilon_k)_{k=1}^T$, where $Y^\epsilon_t$ is the solution of the following SDE:

\begin{equation}
\label{continuous Y}
	dY^\epsilon_t = h(X^{(1,\epsilon)}_t,X^{(2,\epsilon)}_t,Y^\epsilon_t)dt + \tau(X^{(1,\epsilon)}_t,X^{(2,\epsilon)}_t,Y^\epsilon_t) dV_t,\ \ Y_0 = 0.
\end{equation}
Then, we can replace (\ref{continuous Y}) by its averaged approximation
\begin{equation}
\label{averaged Y}
	d\bar{Y}_t = \bar{h}(\bar{X}_t,\bar{Y}_t)dt + \bar{\tau}(\bar{X}_t,\bar{Y}_t) dV_t,\ \ Y_0 = 0,
\end{equation}
for 
\[\bar{h} = \int_{{\mathbb R}^q} h(x,z,y) \nu_x(dz)\ \ \ {\rm and}\ \ \ \bar{\tau} = \left( \int_{{\mathbb R}^q} \tau(x,z,y)^2 \nu_{x}(dz)  \right)^\frac{1}{2}.\]
Notice that, by the averaging principle, $(X^{\epsilon,1}_t,Y^\epsilon_t) \rightarrow (\bar{X}_t,\bar{Y}_t)$, as $\epsilon\rightarrow 0$. Then, we can apply the particle filter described in \cite{DelMoral-Jacod-Protter} for this type of observation process and approximate $\pi^\epsilon_k$ by 
\[ \bar{\pi}_k = {\mathbb P}\left( \bar{X}_k | \bar{Y}_\ell = Y^\epsilon_\ell, \ell=1,\dots,k\right). \]
We expect that the efficiency of the algorithm will also be improved in this case.

In this paper, we introduced a particle filter for the estimation of a quantity ($X^{(1,\epsilon)}_t$) that can be approximated by a diffusion given discrete and partial observations of it, in the case where this quantity is the slow component of a multiscale diffusion of the form (\ref{full X}). The main idea is that rather than evolving the particles by simulating the full multiscale system which can be very inefficient, it is better to do a short runs of of the full multiscale system and use these simulations to locally estimate the drift and variance of the diffusion that approximates the evolution of the partially observed quantity. Depending on the multiscale system and the approximate diffusion, one can use different methods for the estimation of the diffusion parameters and the evolution of the particles that follow the diffusion, rather than the Monte-Carlo estimation and the Euler simulation discussed above. 

\section*{Acknowledgements}
The author would like to thank Professor I.G. Kevrekidis for suggesting this problem to her.


\end{document}